\documentclass[twocolumn,showpacs,showkeys,preprintnumbers,amsmath,amssymb,pra,aps,10pt]{revtex4-2}
\usepackage[T1]{fontenc}
\usepackage{color}
\usepackage{soul}
\usepackage{graphicx}
\usepackage{dcolumn}
\usepackage{bm}
\usepackage[colorlinks=true,linkcolor=blue,citecolor=red]{hyperref}
\usepackage{color}

\begin{document}

\preprint{\emph{Submitted to:} Journal of Magnetism and Magnetic Materials}

\title{Charge-order on the triangular lattice:\\ Effects of next-nearest-neighbor attraction in finite temperatures}%

\author{Konrad J. Kapcia}
\email[\mbox{e-mail: }]{konrad.kapcia@amu.edu.pl}
\homepage[\mbox{ORCID ID: }]{https://orcid.org/0000-0001-8842-1886}
\affiliation{Faculty of Physics, Adam Mickiewicz University in Pozna\'{n},
ulica Uniwersytetu Pozna\'{n}skiego 2, PL-61614 Pozna\'{n}, Poland}

\date{\today}

\begin{abstract}
The extended Hubbard model in the atomic limit, which is equivalent to lattice $S=1/2$ fermionic gas, is considered on the triangular lattice.
The model includes onsite Hubbard $U$ interaction and both nearest-neighbor ($W_{1}$) and next-nearest-neighbor ($W_{2}$) density-density intersite interactions. 
The variational approach treating the $U$ term  exactly and the $W_l$ terms in the mean-field approximation is used to investigate thermodynamics of the model and to find its finite temperature ($T>0$) phase diagrams (as a function of particle concentration) for $W_{1}>0$ and $W_{2}<0$.
Two different types of charge-order (i.e., DCO and TCO phases) within $\sqrt{3} \times \sqrt{3}$ unit cells as well as the nonordered (NO) phase occur on the diagram. 
Moreover, several kinds of phase-separated (PS) states (NO/DCO, DCO/DCO, DCO/TCO, and TCO/TCO) are found to be stable for fixed concentration.
Attractive $W_{2}<0$ stabilizes PS states at $T=0$ and it extends the regions of their occurrence at $T>0$. 
The evolution of the diagrams  with increasing of $|W_{2}|/W_{1}$ is investigated.
It is found that some of the PS states are stable only at $T>0$.
Two different critical values of $|W_{2}|/W_{1}$ are determined for the PS states, in which two ordered phases of the same type (i.e., two domains of the DCO or TCO phase) coexist. 
\end{abstract}

\keywords{charge order, triangular lattice, fermionic lattice gas, longer-range interactions, extended Hubbard model, atomic limit\\
\quad\\
\textbf{Highlights:}\\
\begin{itemize}
\item Atomic limit of the extended Hubbard model on the triangular lattice is analyzed.
\item Phase diagrams of the lattice $S=1/2$ fermionic gas model are found.
\item The effects of next-nearest-neighbor attractive interaction are investigated.
\item The diagrams have complex structure with different multicritical points.
\item The stability regions of various phase separated states are found.
\end{itemize}
}

\maketitle

\section{Introduction}

The classical lattice gas model (equivalent with the $S=1/2$
Ising model) is useful effective model for description of
adsorbed particles on crystalline substrates (cf. pioneering works
on the triangular
lattice~\cite{HoutappelPhys1950A,HoutappelPhys1950B,CampbellPRA1972,KaburagiJJAP1974,MetcalfPhysLettA1974,MihuraPRL1977,Kaburagi1978}).
In the case of a graphine surface  or a single layer of graphene
as well as (111) face-centered cubic surface, the periodic
potential of the underlying crystal surface forms a triangular
lattice, which can be occupied by adsorbed atoms,
e.g.,~\cite{CaragiuJPCM2005,ProfetaPRB2004,Rodriguez2018,Menkah2019,Xing2021}.   
Although the adsorbed particles are rather classical, taking into
account the quantum properties is necessary for a description of
helium atoms
adsorption~\cite{BretzPRL1971,BretzPRA1973,Aziz1989,LitakJMMM2017}.
Thus, in the present work, an extension of the classical lattice gas model to $S=1/2$ fermionic particles is analyzed.
Particular attention is taken for effects of the next-nearest-neighbor attraction on the phase diagrams at finite temperatures.

The investigated model has the form of the extended Hubbard
model~\cite{MicnasRMP1990,GeorgesRMP1996,ImadaRMP1998,DuttaRPP2015,LitakJMMM2017}
in the atomic limit (i.e.,~zero-bandwidth limit)
with Coulomb interactions restricted to the next-nearest neighbors (or, equivalently, to the  second  neighbors) and it can be written as:
\begin{equation}
\label{eq:hamUW} 
 \hat{H}  =  U\sum_i{\hat{n}_{i\uparrow}\hat{n}_{i\downarrow}} + \sum_{l=1,2} \left[ \frac{W_{l}}{2 z_{l}}\sum_{\langle i,j\rangle_{l}}{\hat{n}_{i}\hat{n}_{j}} \right] 
- \mu\sum_{i}{\hat{n}_{i}},
\end{equation}
where $\hat{c}^{\dagger}_{i\sigma}$ ($\hat{c}_{i\sigma}$) is the creation (annihilation) operator of a fermionic particle with spin $\sigma$ ($\sigma \in \{ \uparrow, \downarrow \}$) at the site $i$, whereas $\hat{n}_{i\sigma}=\hat{c}^{\dagger}_{i\sigma}\hat{c}_{i\sigma}$ and $\hat{n}_{i}=\sum_{\sigma}{\hat{n}_{i\sigma}}$ are the number operator of particles with spin $\sigma$ at site $i$ and the total number operator of particles at site $i$, respectively.
$\sum_{\langle i,j\rangle_{l}}$ denotes the summation over $l$th neighbors independently ($l=1,2$). 
$U$ is the onsite interaction, whereas
$W_{1}$ and $W_{2}$ denote the intersite interactions between the nearest neighbors (NNs)
and the next-nearest neighbors (NNNs), respectively.
$z_1$ and $z_2$ are numbers of NNs and NNNs, respectively ($z_{1}=z_{2} = 6$ for the triangular lattice).
The chemical potential $\mu$ is related with the total concentration
$n$ of particles in the system through $n = (1/L)\sum_{i}{\left\langle \hat{n}_{i} \right\rangle}$,
where \mbox{$0\leq n \leq 2$} and $L$ is the total number of lattice sites.

 Within the variational approach treating onsite $U$ term  exactly and intersite $W_{l}$ terms in the mean-field approximation, i.e.,~\begin{equation}
\label{eq:MFAdecoupling}
\hat{n}_{i} \hat{n}_{j} = \langle \hat{n}_{i} \rangle \hat{n}_{j} + \hat{n}_{i} \langle \hat{n}_{j} \rangle - \langle \hat{n}_{i} \rangle \langle \hat{n}_{j} \rangle,
\end{equation}
this model was intensively investigated on the hypercubic
lattices, e.g.,~for
$W_{2}=0$~\cite{BariPRB1971,MicnasPRB1984,Bursill1993} and
$W_{2}\neq0$~\cite{KapciaJPCM2011,KapciaPhysA2016,KapciaJSNM2017,KapciaPRE2017}.
Moreover, rigorous results for one-dimensional chain were found
in~\cite{Macini2008,Mancini2013} as well as the model on
two-dimensional square lattice was analyzed by various
methods~\cite{Borgs1996,Lee2001,Pawlowski2006,Ganzenmuller2008,RademakerPRE2013,KapciaPRE2017}.
On the triangular lattice [within the variational approach with
decoupling (\ref{eq:MFAdecoupling})], the evolution of
metastable phases in the model was determined for $U<0$ and
$W_{2}=0$~\cite{KapciaJSNM2019}, whereas the full phase diagram
(for all temperatures, at $T=0$ and $T>0$) for
$W_{2}=0$ was obtained in~\cite{KapciaNano2021}. 
In~\cite{KapciaNano2021} the effects of $W_{2}<0$ only at the
ground state were also discussed. 

In the present work, the effects of the attractive NNN
interaction (i.e.,~$W_{2}<0$) are investigated at $T>0$
within the variational approach mentioned, for details
cf.~\cite{MicnasPRB1984,Bursill1993,KapciaJPCM2011,KapciaPhysA2016,KapciaPRE2017,KapciaJSNM2019,KapciaNano2021}.
In particular, the evolution of phase diagrams (for fixed $U/W_{T}$, where $W_{T}=W_{1}-2 W_{2}$, $W_1>0$, and $W_2\leq 0$) with increasing $k=| W_{2}| /W_{1}$ is presented and an emergence of novel phase separation states (not occurring for $W_2=0$ or $T=0$) is noticed.
Because the triangular lattice can be divided into three equivalent sublattices only orderings within $\sqrt{3} \times \sqrt{3}$ unit cell are considered (the three-sublattice assumption), which is justified in the range of model parameters considered.  

For $T>0$, the expressions given in~\cite{KapciaJPCM2011}
for the triangular lattice and $W_2\neq 0$ take the following
forms (cf. also these in~\cite{KapciaNano2021,KapciaPRE2017}).
For a grand canonical potential $\omega$ (per lattice site) one obtains
\begin{equation}
\label{eq:grandpotential.fintemp}
\omega = - \frac{1}{6}\sum_{\alpha}\varphi_{\alpha} n_{\alpha} -\frac{1}{3\beta} \sum_{\alpha}  \ln{Z_{\alpha}} , 
\end{equation}
where $\beta=1/(k_{B} T)$ is inverted temperature, coefficients $\varphi_{\alpha}$ are defined as $\varphi_{\alpha}=\mu-\mu_{\alpha}$,
\begin{equation}
\label{eq:Zalpha.fintemp}
Z_{\alpha} = 1 + 2 \exp \left( \beta  \mu_{\alpha} \right) + \exp{\left[ \beta \left( 2\mu_{\alpha} - U \right)\right]},
\end{equation}
and $\mu_{\alpha}$ is a local chemical potential in $\alpha$ sublattice ($\alpha\in\{A,B,C\}$) defined as
\begin{equation}
\label{eq:phi.fintemp}
\mu_{\alpha}  = \mu - \tfrac{1}{2}  W_{1} (n_{\alpha'} + n_{\alpha''}) + W_2 n_{\alpha}.
\end{equation}
Here, $\alpha'$ and $\alpha''$ denote two other sublattices than $\alpha$ (and $\alpha'\neq \alpha''$).
Particle concentration $n_\alpha=(3/L)\sum_{i \in \alpha} \langle \hat{n}_i \rangle$ in each sublattice $\alpha$ for arbitrary $T>0$ is expressed by
\begin{equation}
\label{eq:nalpha.fintemp}
n_{\alpha} = \frac{2}{Z_\alpha} \left\{ \exp{ \left( \beta  \mu_{\alpha} \right) } + 
\exp{\left[ \beta \left( 2\mu_{\alpha} - U \right) \right]} \right\}.
\end{equation}
Three equations for $n_{\alpha}$ determine the solution for a (homogeneous) phase occurring in the system for fixed $U$, $W_{1}$, $W_{2}$, and $\mu$. 
If $n = (n_{A} + n_{B} + n_{C})/3$ is fixed, the set is solved with respect to $\mu$, $n_{A}$, and $n_{B}$ (the third concentration is found as $n_{C}=3n - n_{A} - n_{B}$). 
This set has usually several solution, thus it is extremely important to find the solution corresponding to the lowest $\omega$ (if $\mu$ is fixed) or free energy $f=\omega+\mu n$ (if $n$ is fixed). 
For fixed $n$, the phase-separated (PS) states can also
occur, which free energy is determined by the Maxwell's
construction (macroscopic phase separation),
e.g.,~\cite{KincaidPR1975,ArrigoniPRB1991,BakAPPA2004,Bursill1993,KapciaNano2021,KaganPR2021}.

\section{Numerical results ($W_{1} > 0$ and $W_{2} \leq 0 $)}

It was shown that in the model (within the approximation used)  the following phases can occur:
(i) the nonordered (NO) phase with $n_{A}=n_{B}=n_{C}$,
(ii) the charge-ordered phase with two different concentration in sublattices (the DCO phase, e.g.,~$n_{A} \neq n_{B} = n_{C}$ and other cyclic permutations, $3$ equivalent solutions), and
(iii) the charge-ordered phase with $n_{A} \neq n_{B}$, $n_{B} \neq n_{C}$, and
$n_{A} \neq n_{C}$ (the TCO phase; three different concentrations in
sublattices, $6$ equivalent
solutions)~\cite{KenekoPRB2018,KapciaNano2021}.
Moreover, for $W_2=0$, two PS states were found in some ranges of $n$: 
(i) PS1:NO/DCO, where the NO and the DCO phases coexist and (ii)
PS2:DCO/DCO, where two different DCO phases
coexist~\cite{KapciaJSNM2019,KapciaNano2021}. 
Due to the particle--hole symmetry of model (\ref{eq:hamUW}) the
phase diagram is symmetric with respect to half-filling, i.e.,~$n=1$ or $\bar{\mu}=0$
($\bar{\mu}=\mu - U/2 - W_{1} - W_{2}$)~\cite{MicnasRMP1990,MicnasPRB1984,KapciaJSNM2019,KapciaNano2021}.

\begin{figure}
\includegraphics[width=\linewidth]{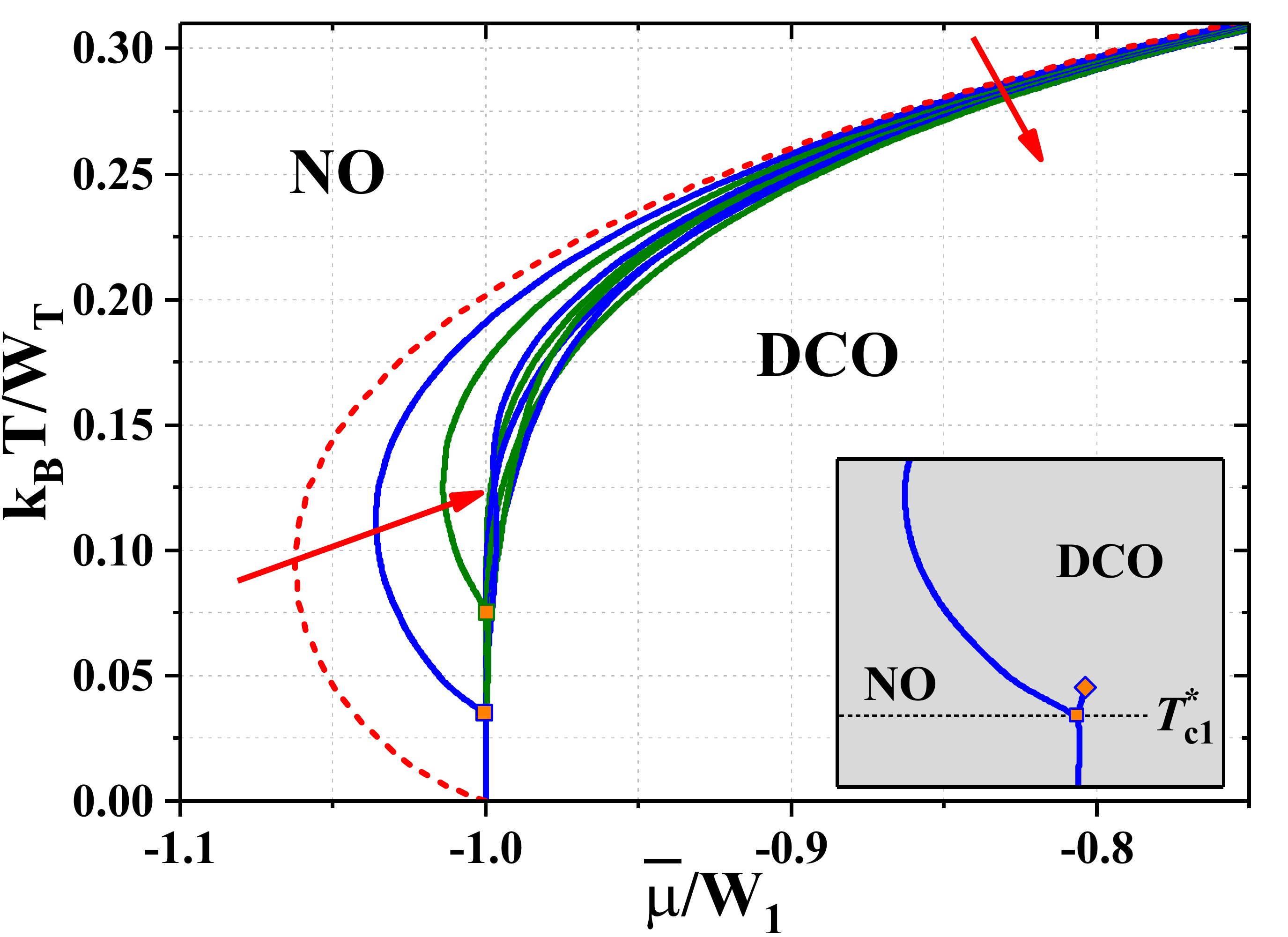}
\caption{The DCO-NO boundary as a function of $\bar{\mu}/W_{1}$ ($\bar{\mu}=\mu - U/2 - W_{1}-W_{2}$) 
 for $U/W_{T}=-1.00$ ($W_{T}=W_{1}-2W_{2}$) and 
 several values of $k=| W_{2}| /W_{1}$ ($k=0.00,0.05,0.10,0.15,0.20,0.25,0.50,1.00,2.00$) 
 shown for $\bar{\mu}/W_{1}<-0.75$.
 The arrows show the direction of increasing $k$. 
 All boundaries are first order.
 At $\bar{\mu}=0$ all lines go to the same value of $k_{B} T_{M}/W_{T}$.
 The inset shows schematically the structure of the diagram with 
 triple and bicritical-end points (denoted by squares and diamonds, respectively) for $0<k<k_{c1}$, 
 where a discontinuous DCO-DCO boundary appears near $\bar{\mu}/W_{1}\approx -1$ at $T>T_{c1}^{*}$. 
 The bicritical-end point and the DCO-DCO line are shown only in the inset.\label{FIG:1}}
\end{figure}

\begin{figure}
\includegraphics[width=\linewidth]{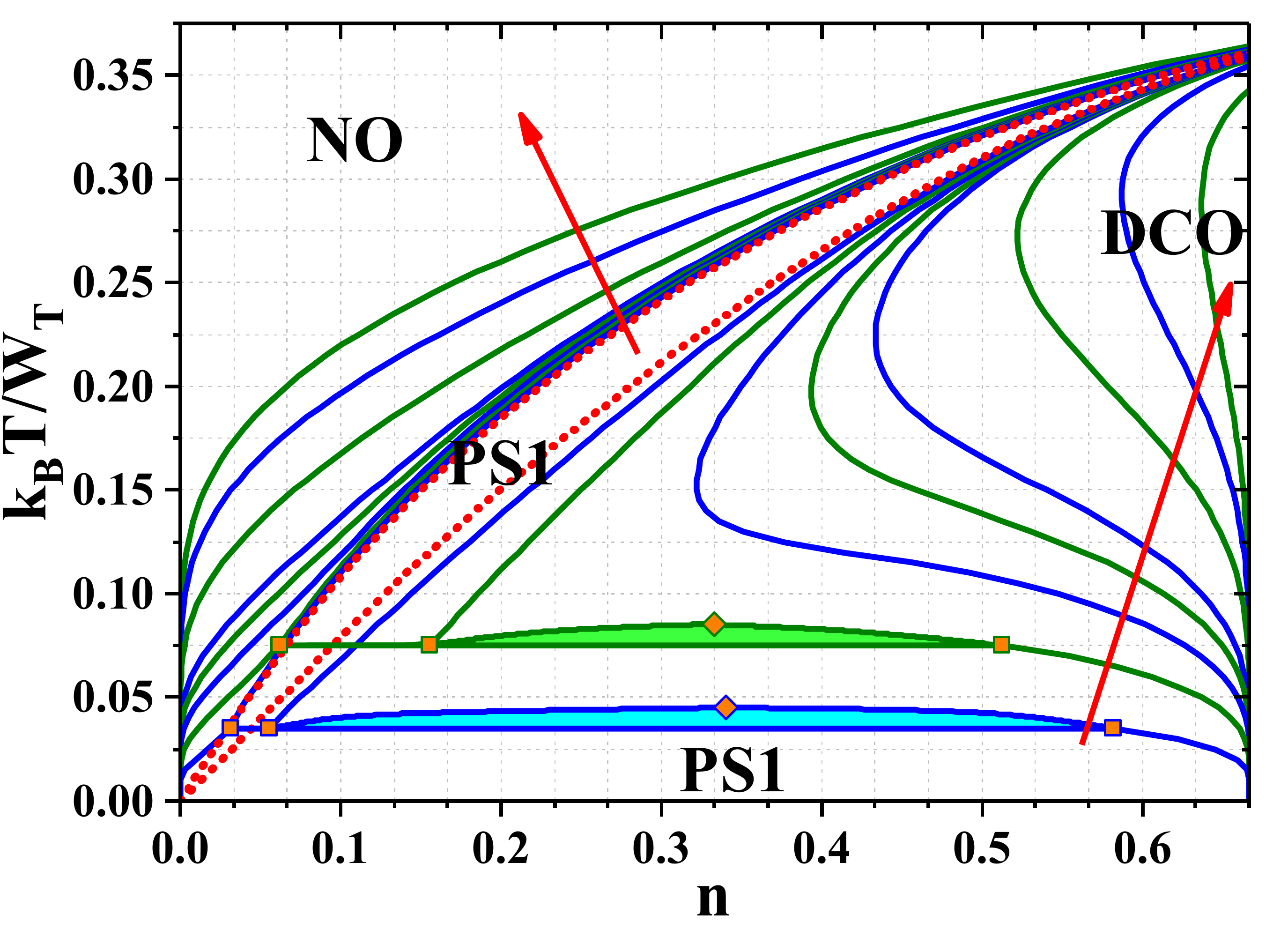}
\caption{Phase diagrams in the range of $0 \leq n \leq 2/3$ for $U/W_{T} = -1.00$ 
 and several values of $k = | W_{2}| /W_{1}$
 ($k=0.00,0.05,0.10,0.15,0.20,0.25,0.50,1.00,2.00$). 
 The arrows show the direction of increasing $k$. 
 The shadowed regions indicate the PS2 state occurrence. 
 The boundaries for $W_{2}=0$ are denoted by dotted lines.
 Symbols denote critical points as in  Fig. \ref{FIG:1}: 
 triple points correspond to three different concentrations.\label{FIG:1A}}
\end{figure}

\begin{figure}
\includegraphics[width=\linewidth]{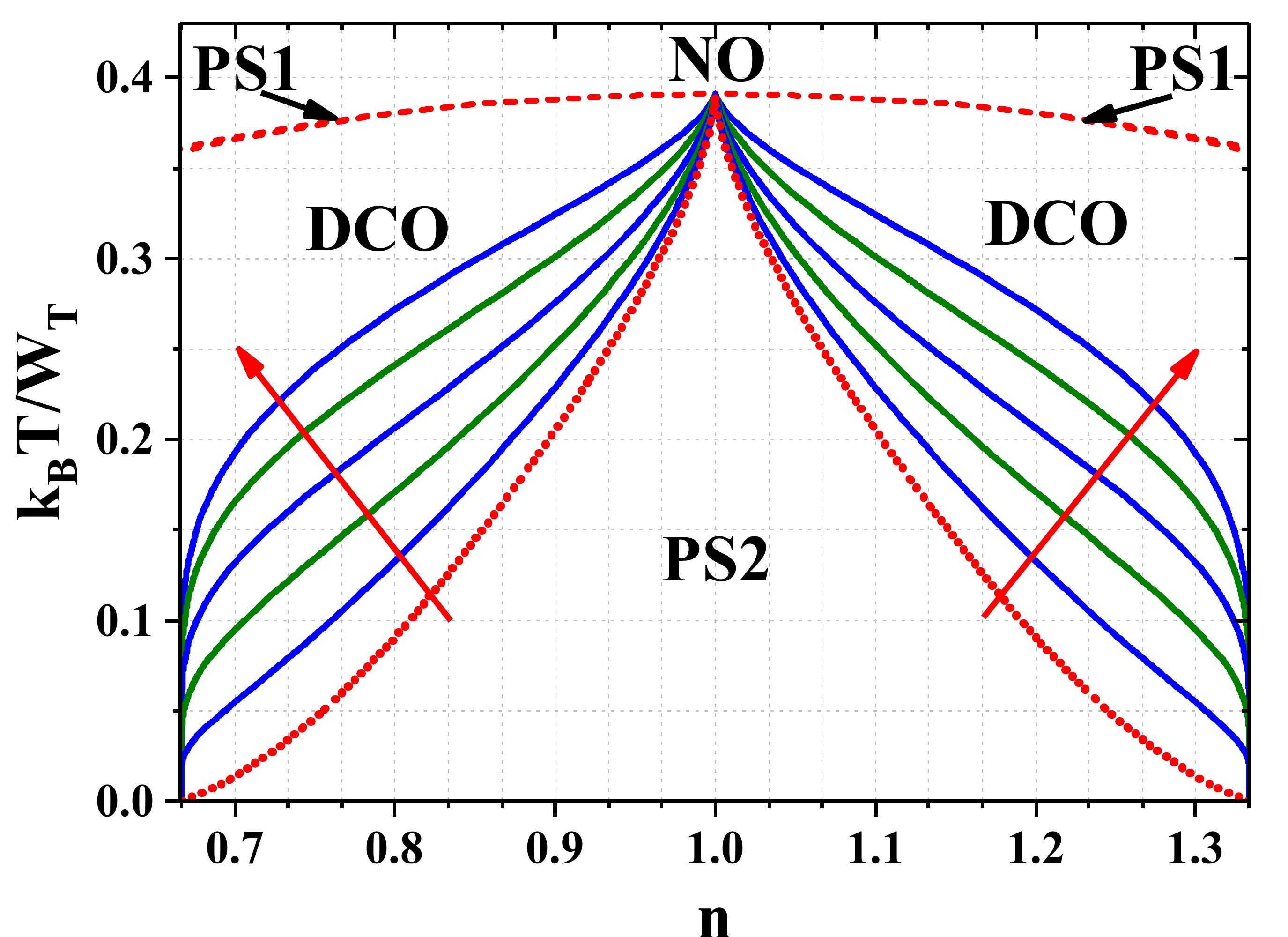}
\caption{Effects of $W_2$ on the PS2 region ($2/3 < n < 4/3$) for $U/W_{T}=-1.00$ 
 and several values of $k=0.00,0.10,0.25,0.50,1.00,2.00$.
 The arrows show the direction of increasing $k$.
 The boundaries for $W_{2}=0$ are denoted by dotted lines.
 Very narrow PS1 regions (dashed lines) at high temperatures are shown only for $k=0$.\label{FIG:2}}
\end{figure}

The simplest phase diagram of the model is for $U < 0$. 
For $W_{2} = 0$, the diagram for fixed $\mu$ consists of two regions of the DCO phase and one region of the NO phase. 
These regions are separated by two kinds of first-order (discontinuous) boundaries: 
(i) the DCO-NO line with its maximum temperature $T_{M}$ at half-filling and 
(ii) the DCO-DCO line at $\bar{\mu}=0$ extending from $T=0$
to $T=T_{M}$ (it is
$\bar{\mu}$-independent)~\cite{KapciaJSNM2019,KapciaNano2021}
(cf. also  Fig. \ref{FIG:1}).
As a result, on the diagram as a function of $n$ one finds regions of two PS states occurrence:
(i) the PS1 state occurring in narrow range of $n$ for $0 < T < T_{M}$ (it does not exist at $T=0$,  Fig. \ref{FIG:1A}) and
(ii) the PS2 state, which is stable for $0 \leq T < T_{M}$ (with concentrations in coexisting domains as $n_{-} = 2/3$ and $n_{+} = 4/3$ at $T = 0$;  Fig. \ref{FIG:2}), respectively.

Nonzero $W_{2} < 0$ extends the regions of the PS states occurrence.
For $W_{2} < 0$, the PS1 state is stabilized at $T = 0$ with $n_{-} = 0$ and $n_{+} = 2/3$. 
For $0 < k < k_{c1}$ (where $k_{c1} \approx 3/20$) and above some $T^{*}_{c1}$ (which is $U/W_T$- and $k$- dependent), the PS2 state appears in a narrow region (shadowed regions in  Fig. \ref{FIG:1A}).
It vanishes continuously at $k = k_{c1}$.
With increasing $T$ for fixed $n$ (higher concentrations), at $T_{c1}^{*}$ the PS1-PS2 transition occurs, which is associated with the change of the phase in the domain of lower concentration (from the NO to the DCO phase) with simultaneous a discontinuous change of concentration $n_{-}$ in this domain. 
For lower concentrations $n$, there is also a transition between two different PS1 states (the PS1-PS1 transition).
At $T^{*}_{c1}$ the discontinuous change of concentration in the DCO phase domain of the PS1 state occurs.  
This behavior is associated with a new first-order DCO-DCO
boundary inside the DCO region [ending at a bicritical-end (also
called as isolated-critical) point,
cf.~\cite{MicnasPRB1984,KapciaJPCM2011,KapciaPhysA2016,KincaidPR1975}],
which is present on the diagram for fixed $\bar{\mu}$. 
This is schematically shown only in the inset of  Fig. \ref{FIG:1}, where three first-order lines merge in the triple point located at $T_{c1}^{*}$.
$T_{c1}^{*}$ increases with $k$ and for $k\rightarrow k_{c1}$ the bicritical-end point goes to the DCO-DCO line.
On the $k_{B}T/W_{T}$--$\bar{\mu}/W_{1}$ diagrams the DCO region shrinks with increasing $k$ and simultaneously the re-entrant feature of the DCO-DCO line is destroyed for large $k$.
However, the evolution of the DCO-DCO boundary is nonmonotonous (these for $k>k_{c1}$ can cross each other, approximately, in the range of $-1.0 < \bar{\mu}/W_1 < -0.9$,  Fig. \ref{FIG:1}).
For $k>k_{c1}$, the PS2 region is absent (the triple and the bicritical-end points do not exist) and only the PS1 state occurs on the diagram for $n<2/3$ (the PS1 region separates the DCO and the NO regions for all $T<T_{M}$).

The evolution of the PS2 region boundaries in the range  $2/3<n<4/3$ is shown in  Fig. \ref{FIG:2}.
Increasing $k$ enlarges the PS2 region existing for $0\leq T < T_{M}$ with simultaneous change of boundary curvatures.

The above discussed behaviors are generic for all $U<| W_{2}| $ (or $U/W_{1} - k<0$).  
For $U/W_{T}>1$ the critical behaviors are similar to those
discussed above, but the structure of phase diagrams exhibits two
lobs of the DCO phase occurrence~\cite{KapciaNano2021} and the
maximal temperature $T_M$ for the DCO-DCO transition is
located for $\bar{\mu}$ corresponding to $n=1/2$ and
$n=3/2$ (cf.
also~\cite{MicnasPRB1984,KapciaJPCM2011,KapciaPhysA2016}).

\begin{figure}
\includegraphics[width=\linewidth]{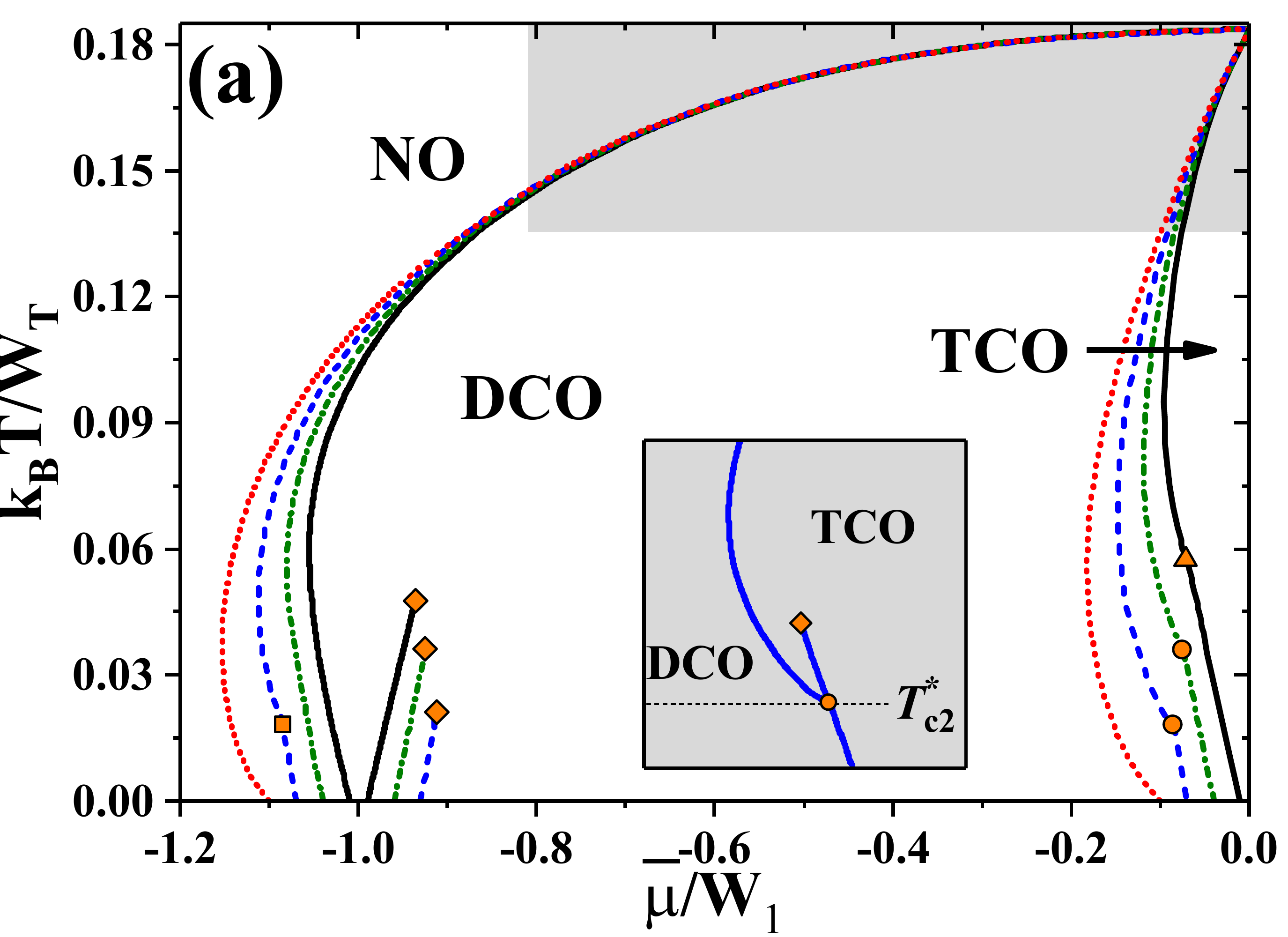}
\includegraphics[width=\linewidth]{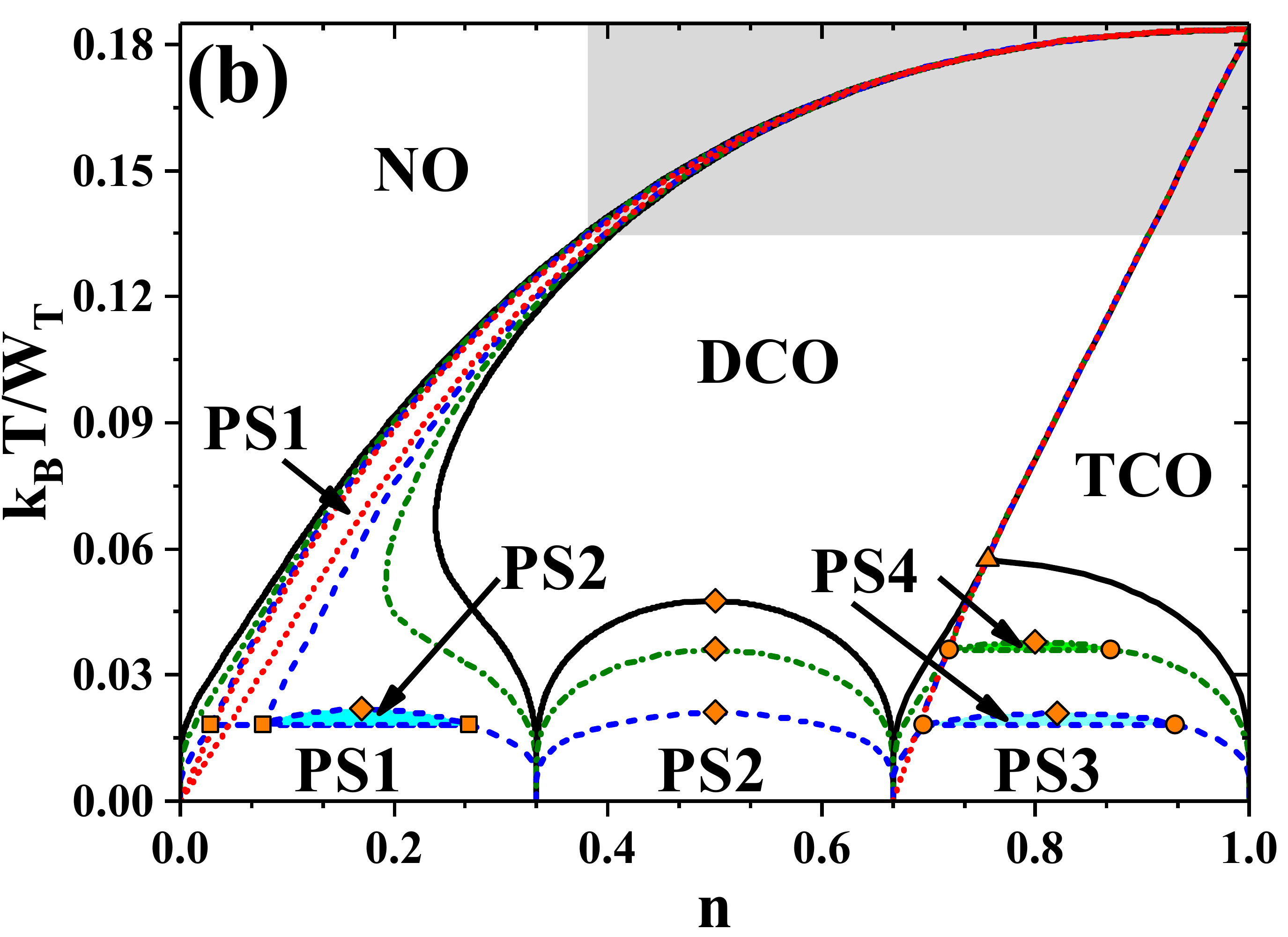}
\caption{The diagram as a function of $\bar{\mu}/W_1$ (a) 
 and as a function of $n$ (b) for $U/W_{T}=0.2$ and several values of $k=0.00,0.10,0.20,0.30$ 
 (dotted, dashed, dash-dotted, and solid lines, respectively).
 The inset of panel (a) shows schematically the structure of the diagram with 
 critical-end and bicritical-end points for $0<k<k_{c2}$, 
 where a discontinuous TCO-TCO line appears at $T>T_{c2}^{*}$.
 Squares, diamonds, circles, and triangles indicate triple, bicritical-end, critical-end, 
 and higher-order critical points.
 Triple and critical-end points correspond to three or two different concentrations, respectively.
 Above critical-end  or higher-order critical points, or for $k=0$, the DCO-TCO transition is second order.
 All other boundaries are first order.
 Not all bicritical-end points, not all DCO-DCO lines, neither no TCO-TCO lines are shown on panel (a).\label{FIG:3}}

\end{figure}

For $U>| W_{2}| $ and $U/W_{T}<1/2$ ($0<U/W_{1}-k<1/2$), the TCO phase appears on the phase diagram (Fig. \ref{FIG:3}).
For $k=0$, the TCO phase is stable near half-filling (in a range of $2/3<n<4/3$ at $T=0$) and
the TCO-DCO transition is continuous (second
order)~\cite{KapciaNano2021}, cf. also Appendix \ref{sec:appendix}. 
For $k>0$ the following qualitative changes occur, which are shown in  Fig. \ref{FIG:3} for $U/W_{T}=0.2$.
The evolution  of the DCO-NO boundary for fixed $\bar{\mu}$ is the same as for the case discussed previously.
With increasing $k$ (for $0<k<k_{c1}$) the triple point appears [with the DCO-DCO boundary at $T>T_{c1}^{*}$, not shown in  Fig. \ref{FIG:3}(a); cf., inset of  Fig. \ref{FIG:1}]  and the PS1 and PS2 states are present in define ranges of $n$, whereas for $k>k_{c1}$ only the PS1 state occurs (but now, at $T=0$, the concentrations in domains of the PS1 state are $n_{-}=0$ and $n_{+}=1/3$).
For smaller $| \bar{\mu}| $ other discontinuous DCO-DCO boundary extending from $T=0$ exists [almost straight lines in  Fig. \ref{FIG:3}(a)], which results in the PS2 state occurrence in $1/3<n<2/3$ range (this behavior is not present for $k=0$). 

The attractive $W_{2} \neq 0 $ affects also on the DCO-TCO boundary.
It changes its order for small $T$ and for $T<T^{*}_{c2}$ the DCO-TCO transition is first order (for fixed $\mu$).
For $T<T^{*}_{c2}$ the PS3:DCO/TCO state, which is a coexistence of the DCO and the TCO phases, is stable (at $T=0$ in the range of $2/3<n<1$).
For $0<k<k_{c2}$ (where $k_{c2} \approx 3/10$) and $T>T^{*}_{c2}$, the PS4:TCO/TCO state appears in a narrow region [ Fig. \ref{FIG:3}(b)], which shrinks with increasing $k$ and vanishes continuously at $k=k_{c2}$.
This behavior is connected with the occurrence of a first-order TCO-TCO transition at $T>T^{*}_{c2}$ [not shown in  Fig. \ref{FIG:3}(a)], which ends at a bicritical-end point [the discontinuous DCO-TCO, the continuous DCO-TCO, and discontinuous TCO-TCO lines merge at  a critical-end point, schematically shown only in the inset of  Fig. \ref{FIG:3}(a)].  
Note that at $T^{*}_{c2}$ the PS3-PS4 transition occurs for fixed $n$, which is associated with the TCO-DCO transition in one domain.
$T_{c2}^{*}$ increases with $k$ and for $k=k_{c2}$ the
bicritical-end and the critical-end points merges into one
critical point of higher order (cf.~\cite{KincaidPR1975}).
For $k>k_{c2}$, the discontinuous TCO-TCO line is no longer present on the diagram and a tricritical point appears, at which the DCO-TCO boundary changes its order.   
As a result, only the PS3 state occurs for $2/3<n<1$ separating the DCO and the TCO regions below the tricritical point.
The discussed behavior is similar to those occurring at the
boundaries between checker-board charge-ordered phase and the NO
phase for $k^{H}_{c}=3/5$ for model (\ref{eq:hamUW}) considered on
the hypercubic lattices~\cite{KapciaJPCM2011,KapciaPhysA2016}.
Note also that for $U/W_{T}=0.2$ and $k>1/3$ one gets that $U/| W_{2}| <1$ and the TCO phase is no longer present on the diagram, which is  the regime discussed at the beginning of this section (cf.  Figs. \ref{FIG:1}--\ref{FIG:2}).
Thus, the tricritical point for the PS3 state exists  for $U/W_{T}=0.2$ only in the range $k_{c2}<k<1/3$. 

One should note that, for $(1/3) \ln (2) <U/W_{T} <1/2$, the phase diagram at high temperatures near half-filling is different than that shown in  Fig. \ref{FIG:3} (these regions are schematically indicated by gray rectangles).
For example, for $k=0$, the maximum of the DCO-NO
transition (for fixed $\bar{\mu}$) is not located at
half-filling, discontinuous DCO-DCO transitions appears at
$T>0$ (that results in new regions of the PS1 and PS2
states stability for fixed $n$), the direct
discontinuous TCO-NO transitions is present in define range of
$\bar{\mu}/W_{1}$ and $n$~\cite{KapciaNano2021}.  
The detailed analysis of these issues in the presence of $W_{2} \neq 0$ is beyond the scope of this work.
However, the main results for the evolution of the DCO-NO
boundaries (with vanishing of the triple point at $k=k_{c1}$
and associated PS1-PS2 transition) and of the TCO-DCO boundary
(with a change of the critical-end point into  the tricritical
point at $k=k_{c2}$ and associated PS3-PS4 transition) are
still valid (cf.
also~\cite{MicnasPRB1984,KapciaJPCM2011,KapciaPhysA2016}).  

\section{Conclusions}

In this work, the extended Hubbard model in the atomic limit [Eq.~(\ref{eq:hamUW}) with $W_{1}>0$ and $W_{2} \leq 0$] on the triangular lattice was investigated.
In particular, the effects of next-nearest-neighbor attraction $W_{2}$ were analyzed in detail.
Increasing $k=| W_{2}| /W_{1}>0$ affects the boundaries by increasing discontinuity of $n$ at the transitions (for fixed $\bar{\mu}$) and  extends the regions of phase separation occurrence for fixed $n$ (at $T=0$ they are stable for any incommensurate fillings, i.e.,~$n\neq 2i/3$, $i=1,2,3$ for  $U/W_{1}-k<0$ and $n\neq j/3$, $j=1,\ldots,6$ for  $U/W_{1}-k>0$).
Two different critical values of $k$ exist, below which at $T>0$ the first-order transition between two phases of the same type  occurs (and corresponding phase separated states for define ranges of concentrations are stable), namely:
(i) $k_{c1}\approx 3/20$ for the DCO-NO line and emerging DCO-DCO line and the PS2 region at $T>T^{*}_{c1}$,
(ii) $k_{c2}\approx 3/10$ for the TCO-DCO boundaries and emerging TCO-TCO line and the PS4 region at $T>T^{*}_{c2}$.
For $k>0$ also the PS2 state is stable for some $T \geq 0$ inside the range of $1/3<n<2/3$ (for $U/W_{2}>1$ and $U/W_{T}<1/2$).

Note that, in the $U\rightarrow - \infty$ limit, model (\ref{eq:hamUW}) reduces to the well-known $S=1/2$ Ising model. 
In the general case, model (\ref{eq:hamUW}) can be mapped onto
the $S=1$ Blume--Capel model in the field with an
effective temperature-dependent single-ion
anisotropy~\cite{MicnasPRB1984,Pawlowski2006}. 

Decoupling (\ref{eq:MFAdecoupling}) of the intersite terms is
exact only in $z_{l}\rightarrow + \infty$ limit~\cite{Muller1989,Pearce1978}.
Thus, it is an approximation for the triangular lattice in the general case.  
It overestimates the critical temperatures and stability regions
of ordered phases and could not properly describe the purely
two-dimensional system investigated, where
Berezinskii--Kosterlitz--Thouless-like state
exists~\cite{KenekoPRB2018}.
However, it is rigorous theory at the ground state for the model considered.
The longer-range interaction or small interactions between
two-dimensional layers can stabilize long-range
order~\cite{MihuraPRL1977,Pearce1975}.  

Hamiltonian (\ref{eq:hamUW}) is a relatively simple toy model and it is oversimplified in many aspects for description of real materials.
However, it can be treated as a benchmark for various approximate approaches for models with finite intersite hopping.  
One should underline here, that in the case of nonzero hopping,
for $U<0$ (and any $W_{l}$) or $U>0$ and
$W_{l}<0$, various superconducting states could appear and the
stability ranges of the charge-ordered phases might be reduced
(cf., e.g.,~the results for hypercubic
lattices~\cite{MicnasRMP1990,RobaszkiewiczPRB1981,RobaszkiewiczPRB1982,OlesPLA1984,MicnasJPC1988,MicnasPRB1988}).
Moreover, an occurrence of the Moir\'e pattern (which is the
triangular lattice with a very large supercell) in the
twisted-bilayer graphene (associated with emergence of
superconductivity)~\cite{CaoSci2018,YankowitzScience1059} and
hetero-bilayer transition metal
dichalcogenides~\cite{Wang2018,Xu2020} makes further studies of
various models on the triangular lattice worthwhile.
Also ultra-cold atomic gases on the triangular lattice created by
laser
trapping~\cite{BeckerNJP2010,StruckScience2011,GeorgescuRMP2014,DuttaRPP2015}
are systems, which could enable testing of some theoretical
predictions of this work. 

\subsection*{Acknowledgments}

The author thanks R. Micnas for very useful discussions on some issues raised in this work.
The support from the National Science Centre (Poland) under Grant SONATINA 1 no. UMO-2017/24/C/ST3/00276 is acknowledged. 
Founding in the frame of a scholarship of the Minister of Science and Higher Education (Poland) for outstanding young scientists (no. 821/STYP/14/2019) is also appreciated.

\subsection*{Declaration of competing interest}

The authors declare that they have no known competing financial
interests or personal relationships that could have appeared to
influence the work reported in this paper.
The funders had no role in the design of the study; in the collection, analyses, or interpretation of data; in the writing of the manuscript, or in the decision to publish the results.

\subsection*{CRediT authorship contribution statement}

Konrad Jerzy Kapcia: Conceptualization, Methodology, Software,
Validation, Formal analysis, Investigation, Resources, Data curation,
Writing – original draft preparation, Writing – review \& editing, Visualization,
Supervision, Project administration, Funding acquisition.

\appendix{}
\section{Equation for continuous boundaries}
\label{sec:appendix}
 
One defines $u=\exp(-\beta U)$, $r_{\alpha}=\exp(\beta \mu_{\alpha})$, and $x_{\alpha}=2(r_{\alpha}+ r_{\alpha}^{2} u)/(1+2r_{\alpha} +r_{\alpha}^{2} u)$ [cf. (\ref{eq:nalpha.fintemp}), $n_{\alpha} = x_{\alpha}$].
Introducing $\Delta=(n_A-n_B)/2$ and $\chi=(n_B-n_C)/2$ in (\ref{eq:phi.fintemp}) one gets 
$\mu_{A} = (\chi + 2 \Delta ) W_{T} / 3 + \mu^{*}$,
$\mu_{B} = (\chi - \Delta ) W_{T}  / 3 + \mu^{*}$, and
$\mu_{C} = (-2 \chi - \Delta ) W_{T}  / 3 + \mu^{*}$, where $\mu^{*}=\mu - n(W_{1}+W_{2})$.
In the limit $\chi \rightarrow 0$ ($n_{B} \rightarrow n_{C}$) equation $(x_{B}-x_{C})/(2\chi)=1$ takes the form  of $y_{B}-y_{C}=2$ (by using de l'Hospital theorem), where $y_{\alpha} = \partial x_{\alpha}/\partial \chi$.
Thus, the equation determining continuous transition temperature $T_{c}$ (at which $n_{B} \rightarrow n_{C}$) is
\begin{equation}
\label{eq:2ndorder}
\frac{k_{B} T_{c}}{W_{T}} = \frac{\bar{r} (1+2\bar{r} \bar{u} +\bar{r}^{2} \bar{u}) }{ (1+2\bar{r} +\bar{r}^{2} \bar{u})^{2} },
\end{equation}
where $\bar{r} = \exp (\beta_{c} \mu_{BC})$, $\mu_{BC} = \mu^{*} - \Delta W_{T} / 3$, $\bar{u} = \exp(-\beta_{c} U)$, and $\beta_{c} = 1/(k_{B} T_{c})$ [formally, the positive solution of (\ref{eq:2ndorder}) for $T_{c}$ can exist only for $W_{T}=W_1-2W_2>0$].
The solutions of (\ref{eq:2ndorder}) with $\Delta \neq 0$ coincide with the second-order TCO-DCO lines presented in  Fig. \ref{FIG:3} (above critical points).
The DCO-NO boundary  also coincides with the solution of  (\ref{eq:2ndorder}) with $\Delta = 0$, but only at $T=T_{M}$ (for $W_{2}=0$, they also agree at $T=0$). 
However, such determined $T_{c}/W_{T}$ for $U/W_{T}$ is two
times smaller than corresponding temperature $T_{c}^{H}/W_{Q}$ of
continuous order--disorder transition for $U^{H}/W_{Q} = 2U/W_{T}$ on
hypercubic lattices ($W_{Q}=W_{1}-W_{2}$) for the same $n$,
cf.~\cite{MicnasPRB1984,KapciaNano2021}. 
Other solutions of (\ref{eq:2ndorder})  (also those with
$\Delta = 0$) correspond to transitions between metastable or
unstable phases being solutions of
(\ref{eq:nalpha.fintemp})~\cite{Kapcia2012,KapciaNano2021}.

\bibliography{PM21literatureOK}

\end{document}